\documentclass[epj]{svjour}
\usepackage{graphics}
\usepackage{color}
\begin{document}
\title{Entropic forces exerted on a rough wall by  a  grafted semiflexible polymer}
\author{Parvin Bayati \and Leila Ghassab \and Ali Najafi% etc
% \thanks is optional - remove next line if not needed
\thanks{\emph{Electronic address:} najafi@znu.ac.ir}%
}                     % Do not remove
%
%\offprints{}          % Insert a name or remove this line
%
\institute{Physics Department, University of Zanjan, Zanjan 45371-38791, Iran}
\date{Received: date / Revised version: date}
% The correct dates will be entered by Springer
%
\abstract{We study the entropic force due to a fluctuating  semiflexible polymer that is grafted from one end and  confined  by a rigid and rough wall from the other end.
We show how roughness of the wall modifies the entropic force. In addition to the perpendicular force that is present in the case of a flat wall,  roughness of
the wall adds a lateral component to the force. Both perpendicular and lateral components of the 
force are examined for different values of amplitude and wavelength of the roughness and at different temperatures.
The lateral force is controlled by the local slope of the wall while the perpendicular force is
only sensitive to the curvature of the wall. We show that for small compression, the entropic force is  increased by increasing the curvature of the confining wall. In addition to the biophysical relevance, the results may also be 
useful in developing  an AFM based experimental technique for probing the roughness of surfaces.  
 % end of PACS codes
} %end of abstract
\maketitle

\section{Introduction}
Statistical properties of polymers are crucially enhanced when they live in a confined space \cite{degennes,edwards,kassner,confinanorod}. In a system that undergoes thermal fluctuations, 
entropy reduction due to the confinement eventually shows itself 
as a force that pushes the boundaries. Soft matter is a field that the track of such entropic forces can be examined. Inspired by biological systems, the statistical mechanics of 
polymers has attracted many interests. Microtubules
and actin filaments are among the  biopolymers with range of  important
functionalities in the mechanics of cells \cite{thecell,howard,science,aliee}. Mechanics of cell division 
is also controlled by the forces exerted on the cell cortex from these biopolymers. Self locomotion of some microorganisms like the bacterial pathogen Listeria, is another example that 
the forces from cytoskeleton actin polymers play important role \cite{listeria1}. In addition to the polymerization forces, the entropic forces also contribute in dynamical processes in all above systems. Proper description of the dynamical processes in such systems needs a comprehensive understanding of 
the physics of entropic forces.

In addition to the above biological systems, another relevant motivation for studying the entropic forces comes from the recent single molecule manipulation techniques \cite{singlemol1,singlemol2,singlemol3,singlemol4}.
These techniques, using atomic force microscopy,  enable detailed study of the surface morphology. In recent experiments, to enhance the precision of the surface measurements,  it is
proposed to study the entropic forces due to  a polymer attached to the tip of AFM from one of its 
ends, interacting with a surface from the other end \cite{kardar1,kardar2}. In quantifying the results of such experiments, a theoretical  knowledge about the entropic forces from a fluctuating polymer 
confined by a wall with
specific morphology is necessary.

In this article we address the problem of entropic forces from a semiflexible polymer that is
grafted from one of its ends. 
The fluctuating end of this polymer is
confined with a wall. The problem of entropic forces for this system has been previously considered \cite{gholami}. In current article, we revisit the same problem by letting the wall to be
rough.  To model the roughness, we consider the case of a weak roughness; a wall that is
slightly deformed around a flat reference wall.

The structure of this article is as follows: In section 2, we define the model and present the main
statistical properties that we are going to calculate. In section 3, we present the mathematical
method which we use and finally, in section 4 we give the quantitative results and discuss them.

\section{Model}
Figure \ref{fig1}, shows a schematic of our problem, one end of a semiflexible polymer is clamped and the fluctuations of the other end
is confined by a rigid and rough wall. The grafted end of the polymer is assumed to be perpendicular to the
confining wall.  The  contour length of this polymer is $L$ and its bending rigidity is $\kappa$.
Denoting the strength of thermal fluctuations by $k_BT$, we can define a length scale as: $\ell_p=\kappa/k_BT$.
This persistence length, is a length scale beyond which the thermal fluctuations wash out the correlations between
the tangent vectors of the polymer. The semiflexible polymer, is a polymer that its persistence length has more or less
the same value as its contour length. To study the statistical mechanics of this system, we start with its Hamiltonian:
\begin{figure}
\centering
\resizebox{0.4\textwidth}{!}{%
  \includegraphics{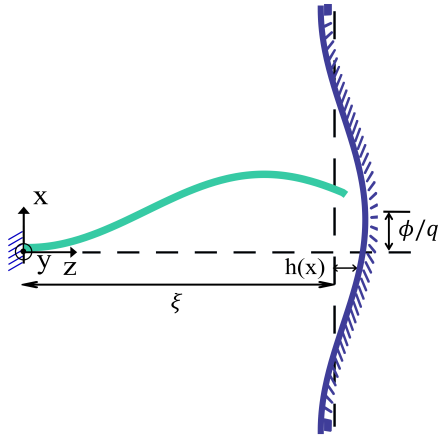}}
\caption{\small A fluctuating semiflexible polymer with one clamped end is shown. A rigid and rough wall confines the
fluctuations of the other end of this polymer.}
\label{fig1}
 \end{figure}
\begin{equation}
\beta{\cal H}=\frac{\ell_p}{2}\int_{0}^{L} ds\left(\frac{d^2}{d s^2}{\bf r}(s)\right)^2+\beta U[{\bf r}(s)],
\end{equation}
where $\beta=1/k_BT$ and ${\bf r}(s)$ is the position vector of a general point on the polymer with contour length $s$ measured from its clamped end.
The first term in this energy functional, is an integral over the local curvature of the polymer and stands for the bending energy of the polymer. The potential energy due to the
presence of  wall is encoded in $U$. As shown in Fig. \ref{fig1}, the coordinate system is chosen in such a way that the clamped end of the polymer is located
at the origin. The local tangent to the polymer at this clamped end is parallel to $z$ axis (perpendicular to the wall). Furthermore the position of the wall is given
by $z=\xi+h(x)$, where
%$\xi$ is the minimum distance between the wall and grafted end of polymer. The 
 roughness of the wall is given by function $h(x)$. We choose an explicit form for the roughness as:
\begin{equation}
h(x)=h_0\cos(q x+\phi),
\end{equation}
where $h_0$ and $2\pi/q$ are the amplitude and wavelength of the roughness and $\phi$ shows how the roughness is symmetric with respect to $x=0$ plane. 
Here we are considering a wall that is curved in $x$ direction and it is flat
in $y$ direction.
Please note
that for small roughness, the case that we are interested in this work, the radius of curvature and 
slope of the wall at point $x=0$ is given by 
$$C=h_0q^2\cos\phi,~~~~S=h_0 q\sin\phi.$$
These two important quantities that are measured just in front of the polymer, 
will be the dominant  characteristics of the roughness that control the entropic force. 

To model the wall potential, we
assume that the interaction between the polymer and the wall is very short range and repulsive. This very short range  interaction energy is given by:
\begin{equation}
U=\left\{\begin{array}{c}
           0 ~~~~~~~~~  r_z(L)<\xi+h(r_x(L))\\
           \infty ~~~~~~~~  r_z(L)\geq \xi+h(r_x(L)),
         \end{array}\right.
\end{equation}
where ${\bf r}(L)$ shows  the position vector for the fluctuating end of polymer.

Our goal in this article is to calculate the entropic forces exerted on the wall.
Denoting the $i$'th component of the force exerted by the polymer on the wall by $f_i(\xi)$, we can calculate it as:
\begin{equation}
  \left\langle f_i(\xi)\right\rangle=\frac{1}{{\cal Z}(\xi)} \int{{\cal D}}[{\bf r}(s)]e^{-\beta({\cal H}_0+U)}\frac{\partial U}{\partial r_i(L)},
  \label{force}
\end{equation}
where ${\cal H}_0$ stands for the Hamiltonian of a freely fluctuating semiflexible polymer and
${\cal Z}(\xi)$ is the partition of the systems and given by:

%%\textcolor{red}{please instead of mathbf{r}, use {bf r} as follows }

\begin{equation}
  {\cal Z}(\xi)=\int {{\cal D}}[{\bf r}(s)]e^{-\beta({\cal H}_0+U)}.
\end{equation}
In evaluating the statistical properties of the system, the functional integrals should be performed  over all configurations of the polymer.
The measure of the functional integral will be chosen in a way  that the partition function 
for a free polymer with $U=0$, normalized to $1$.
For the above prescription  for the interaction potential between the polymer and wall, we can see that:
$$
e^{-\beta U}=\Theta(\xi+h(r_x(L))-r_z(L)),
$$
where $\Theta(x)$, is the Heaviside step function.  In evaluating the partition function of our system we will use this step function representation of the
interaction energy.

As we said before, we are  interested in studying the confinement effects due to a wall that is
slightly deformed around a flat reference plane. In this case, 
the function $h(x)$  that gives the roughness of the  wall, is a small quantity:
$h(x)\ll \xi$. Furthermore the small roughness assumption implies that the roughness wavelength  should 
 also be larger than the polymer tip's undulation: $q r_x(L)\ll 1$.
Now for a wall that is slightly deformed around a reference flat configuration, the interaction potential can be expanded in terms of
$q$. Up to the leading orders of this small quantity we will have:
\begin{eqnarray}
   e^{-\beta U} && \approx\Theta(\xi_r-r_z(L)) \nonumber \\
                && - (S r_x(L) +\frac{1}{2} C r_x(L)^2 )\delta(\xi_r-r_z(L)),
 \end{eqnarray}
where $\xi_r=\xi+h_0\cos\phi$ and $S$ and  $C$ stand for the local slope and curvature of the wall that were previously defined. In this equation, the first term represents the potential due to a flat wall, the second and third
terms show the interaction with an inclined and curved wall respectively.

\section{Mathematical details}
To set up a systematic way for  integrating over the polymer configurations,
we will use the Fourier mode analysis.  Before introducing the Fourier transform, we define the local tangent of the polymer as:
${\bf t}(s)=d{\bf r}(s)/ds$. The condition $|{\bf t}(s)|=1$ would suggest the following
representation for this unit tangent vector:
\begin{equation}
 {\bf t}=\frac{1}{\sqrt{1+a_x^2(s)+a_y^2(s)}}\left(
                                           \begin{array}{c}
                                             a_x(s) \\
                                             a_y(s) \\
                                             1 \\
                                           \end{array}
                                         \right).
\end{equation}
At zero temperature, the polymer is flat and ${\bf t}(s)={\hat z}$. As temperature increases the functions $a_x(s)$ and $a_y(s)$ will start to deviated slightly from zero.
For a semiflexible polymer and in the limit of a weakly bending filament, the deformations are small and we will use a perturbation analysis in terms of small dimensionless functions $a_x$ and $a_y$.
The grafted point of the polymer is always perpendicular to the wall and the other end is free to rotate. Taking into account these conditions we can easily see that the following
boundary conditions should be satisfied by the functions $a_x$ and $a_y$:
$$a_x(0)=a_y(0)=0,~~~~\dot{a}_x(L)=\dot{a}_y(L)=0.$$
Now and to study the deformation, we define the Fourier representation as:
\begin{equation}
 a_x(s)=\sum_{k=1}^\infty a_{x,k}\sin(\lambda_k\frac{s}{L}),~~a_y(s)=\sum_{k=1}^\infty a_{y,k}\sin(\lambda_k\frac{s}{L}).
\end{equation}
Imposing the above boundary conditions, the wavelength of the deformations will be fixed 
as :$\lambda_k=\frac{\pi}{2}(2k-1)$.
In terms of the Fourier modes, the Hamiltonian of a free semiflexible polymer  can be written as:
\begin{equation}
  \beta {\cal H}_0 =\frac{l_p}{4L}\sum_{k=1}^\infty \lambda_k^2(a_{x,k}^2+a_{y,k}^2).
\end{equation}
Now the position vector of the fluctuating end of the polymer can be given as:
\begin{eqnarray}
    r_z(L) && =\int_0^L t_z(s)\approx L-\frac{1}{2}\int_0^L ds \left(a_x^2+a_y^2\right) \nonumber \\
           && =L-\frac{L}{4}\sum_{k=1}^\infty \left(a_{x,k}^2+a_{y,k}^2\right),\nonumber
\label{rz}
\end{eqnarray}
$$
r_x(L)\approx \sum_{k=1}^\infty a_{x,k} \int_0^L ds \sin(\lambda_k
s/L)=L \sum_{k=1}^\infty \lambda_k^{-1}a_{x,k}.
$$
%\begin{equation}
%r_x(L)\approx \sum_{k=1}^\infty a_{x,k} \int_0^L ds \sin(\lambda_k
%s/L)=L \sum_{k=1}^\infty \lambda_k^{-1}a_{x,k},\nonumber
%\label{rx}
%\end{equation}
Please note that these variables are calculated up to the leading orders of small quantities $a_x$ and $a_y$.
In the next section we will use   these results to study the statistical mechanics of the confined semiflexible polymer.
\begin{figure}
\centering
\resizebox{0.5\textwidth}{!}{%
  \includegraphics{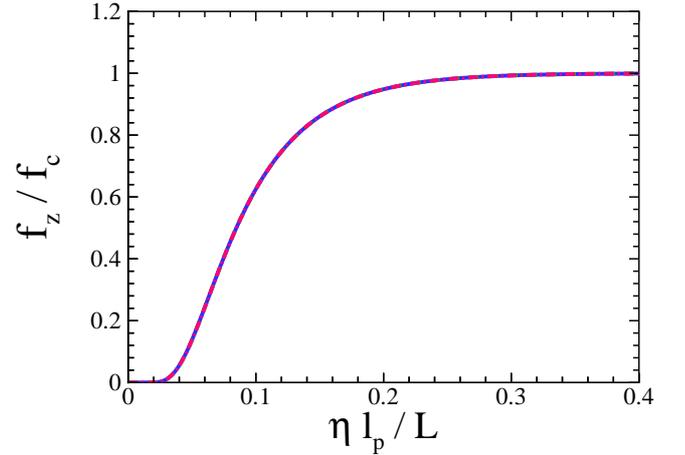}}
%\onefigure[scale=0.3]{02-a.eps}
\caption{For a polymer that is confined by a flat wall from one end 
and clamped at the other end, the entropic
forces have only a component that is perpendicular to the wall. This pushing force is plotted in terms of $\eta\times \ell_p/L$, the longitudinal compression of the polymer. For small compression of the polymer the perpendicular force is starts from zero. The maximum force that can be exerted by a fluctuating polymer is still given by the classical Euler's buckling threshold.
 }
\label{flatwallFzEta}
\end{figure}

Now we are in a position that can proceed and calculate the forces. Regarding the general formula for force (Eq. \ref{force}), we can expand the force in terms of the small quantity $h_0/\xi$. At leading order of the expansion we will have:
\begin{equation}
  \frac{\langle f_z(\xi_r)\rangle}{f_c}=\frac{4L^2/\pi^2l_p}{{\cal Z}_0(\xi_r)}
  \left(\frac{\partial{\cal Z}_0(\xi_r)}{\partial\xi_r}+\frac{\partial{\cal A}(\xi_r)}{\partial \xi_r}-
  {\cal A}\frac{\partial\ln{\cal  Z}_0(\xi_r)}{\partial\xi_r}\right)\nonumber
\end{equation}
\begin{equation}
  \frac{\langle f_x(\xi_r)\rangle}{f_c}=\frac{4L^2}{\pi^2l_p}\frac{{\cal G}(\xi_r)}{{\cal Z}_0(\xi_r)}\nonumber
\end{equation}
where
${\cal Z}_0(\xi_r)$ is the partition function for a fluctuating polymer confined by a flat wall and 
the classical Euler's buckling threshold $f_c=\frac{\pi^2\kappa}{4L^2}$, is used to make the 
components of force nondimensional. The buckling force, measures
the maximum tangential force that a flexible rod (not fluctuating) can tolerate without bending, beyond this force the rod is unstable and it starts to buckle \cite{landau}. 
We will see that the maximum value for the entropic force is given by this buckling threshold.
One should note that this buckling threshold  critically depends on the boundary conditions that we have applied to the rod.
The partition function for a polymer confined by a flat wall is written as:
\begin{equation}
  {\cal Z}_0(\xi_r)= \int{\cal D}[{\bf r}(s)]\Theta\left(\xi_r-r_z(L)\right)e^{-\beta {\cal H}_0},
\end{equation}
and functions ${\cal A}$ and ${\cal G}$ that make the roughness contribution are given by:
\begin{eqnarray}
  {\cal A}(\xi_r)    && =-\int{\cal D}[{\bf r}(s)]\left(S r_x(L)  +\frac{1}{2} C r_x(L)^2\right)
  \delta(\xi_r-r_z(L))e^{-\beta {\cal H}_0},\nonumber\\
  {\cal G}(\xi_r)    && =-\int{\cal D}[{\bf r}(s)]\frac{\partial \left(S r_x(L)
  +\frac{1}{2} C r_x(L)^2\right)}{\partial r_x(L)}\delta\left(\xi_r-r_z(L)\right)e^{-\beta {\cal H}_0}.\nonumber
\end{eqnarray}
Let us now start by evaluating the function ${\cal A}$. We first define the Laplace's transform as:
$$A(f)=\int_{-\infty}^\infty e^{-\beta f \xi}{\cal A}(\xi)d\xi.$$
Performing the integration we will have:
\begin{equation}
  A(f) =  -\int {\cal D}[{\bf r}(s)]e^{-\beta({\cal H}_0+f (L-r_z(L)))}\left(S r_x(L)
                    +\frac{1}{2} C r_x(L)^2 \right),
\end{equation}
To perform the summation over the polymer configurations we will use the Fourier mode analysis.
Up to the leading order of the deformations, all integrals are Gaussian and can be analytically done.
We will have:
\begin{eqnarray}
  A(f) && =-\frac{1}{2} C \int\prod_{i=1}^\infty\frac{da_{x,i}da_{y,i}}{\mathcal{N}_i}\nonumber \\
  &&\exp\left[-\frac{1}{4}\left(\frac{l_p}{L}\lambda_i^2+\beta f L\right)a^{2}_{i}\right]
   \sum_{k=1}^\infty L^2\lambda_k^{-2}a_{x,k}^2 = \nonumber\\
   &&-\frac{1}{2} C \sum_{k=1}^\infty\frac{L^3}{l_p \lambda_k^4}\left(1+\frac{\beta f L^2}{\lambda_k^2 l_p}\right)^{-1}
   \prod_{i=1}^\infty\left(1+\frac{\beta f L^2}{\lambda_i^2 l_p}\right)^{-1},\nonumber
\end{eqnarray}
where $a^{2}_{i}=\left(a_{x,i}^2+a_{y,i}^2\right)$ and ${\cal N}_i = \frac{4 \pi L}{\lambda_i^2 l_p}$.
As one can see, the term proportional to the local slope $S$ has no contribution here. 
The function ${\cal G}$, that makes the lateral components of the force,  
will have a contribution from $S$.  
Now as $A(f)$ is obtained, we can use the definition of the
inverse Laplace transform as:
%$$\frac{1}{2\pi i}\lim_{T\rightarrow\infty}{\cal A}(\xi)=\int_{-\gamma+i T}^{\gamma+i T} 
%\beta e^{\beta f\xi}A(f)df,
%$$
\begin{equation}
  {\cal A}(\xi_r)=\int_{-i\infty}^{i\infty}\frac{\beta}{2\pi i}df e^{\beta f(L-\xi_r)}A(f),
\end{equation}
that can be used to  calculate the required function ${\cal A}(\xi_r)$. 
Now defining $\eta = 1-\xi _r/L$, we can write the final results as:
\begin{eqnarray}
  {\cal A}(\xi_r) && = 2\ C L\left[
         \sum_{k=1}^\infty e^{-\eta \lambda_k^2 l_p/L}
         \prod_{m\neq k}\left(1-\frac{\lambda_k^2}{\lambda_m^2}\right)^{-1}\right.\nonumber\\
    && \times \left(-\frac{l_p}{L}\eta +\sum_{l\neq k}\lambda_l^{-2}\left(1-\frac{\lambda_k^2}{\lambda_l^2}\right)^{-1}\right)\nonumber \\
    && - \sum_{k=1}^\infty \lambda_k^{-4}\sum_{l\neq k}\lambda_l^{2}\ e^{-\eta \lambda_l^2 l_p/L} \left(1-\frac{\lambda_l^2}{\lambda_k^2}\right)^{-2}\nonumber \\
    &&\left.\times\prod_{m\neq k,l}\left(1-\frac{\lambda_k^2}{\lambda_m^2}\right)^{-1}\right].
\end{eqnarray}
Performing a similar steps of calculations one can calculate two important function ${\cal G}$ and 
${\cal Z}_0$. The results read:
\begin{equation}
  {\cal G}(\xi_r)= 2 S \frac{l_p}{L^2}\sum_{k=1}^\infty (-1)^{k+1}\lambda_k e^{-\eta \lambda_k^2 l_p/L},
\end{equation}
\begin{equation}
  {\cal Z}_0(\eta)=2\sum_{k=1}^\infty(-1)^{k+1}\lambda_k^{-1}e^{-\eta \lambda_k^2 l_p/L}.
\end{equation}
For very small values of $\eta$, the above summation over wave vectors in ${\cal Z}_0$ does not rapidly converge. To overcome this convergence issue, we can use the method described in 
reference \cite{gholami} to  achieve a rapidly convergent  representation for ${\cal Z}_0$. 
Following the method presented in above reference, the partition function reads:
\begin{equation}
  {\cal Z}_0(\eta) = 1 + 2 \sum_{l=1}^\infty (-1)^{l}erfc\left(\frac{2 l - 1}{2 \sqrt{\eta l_p/L}}\right)
\label{zconvergent}
\end{equation}
where the error function is given by: 
$$erfc(x)=1-\frac{2}{\sqrt{\pi}}\int_{0}^{x}e^{-t^2}dt.$$

\begin{figure}
\centering
\resizebox{0.5\textwidth}{!}{%
 \includegraphics{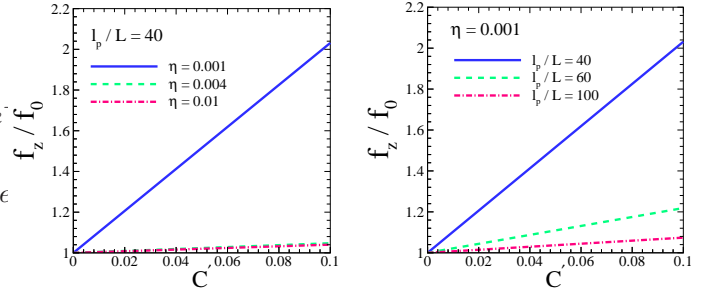}}
  %\includegraphics[width=.8\columnwidth]{03-a-F(x).eps}
%\onefigure[scale=0.3]{03-a-F(x).eps}
\caption{Perpendicular component of the entropic forces depend on the curvature of wall.
For very small compression ($\eta$), the force is plotted in terms of dimensionless  curvature $C'=CL$ and for different values
of persistence length and $\eta$. Here $f_0$, is the value of force for a flat wall.}
\label{roughfz}
\end{figure}

\begin{figure}
\centering
\resizebox{0.5\textwidth}{!}{%
  \includegraphics{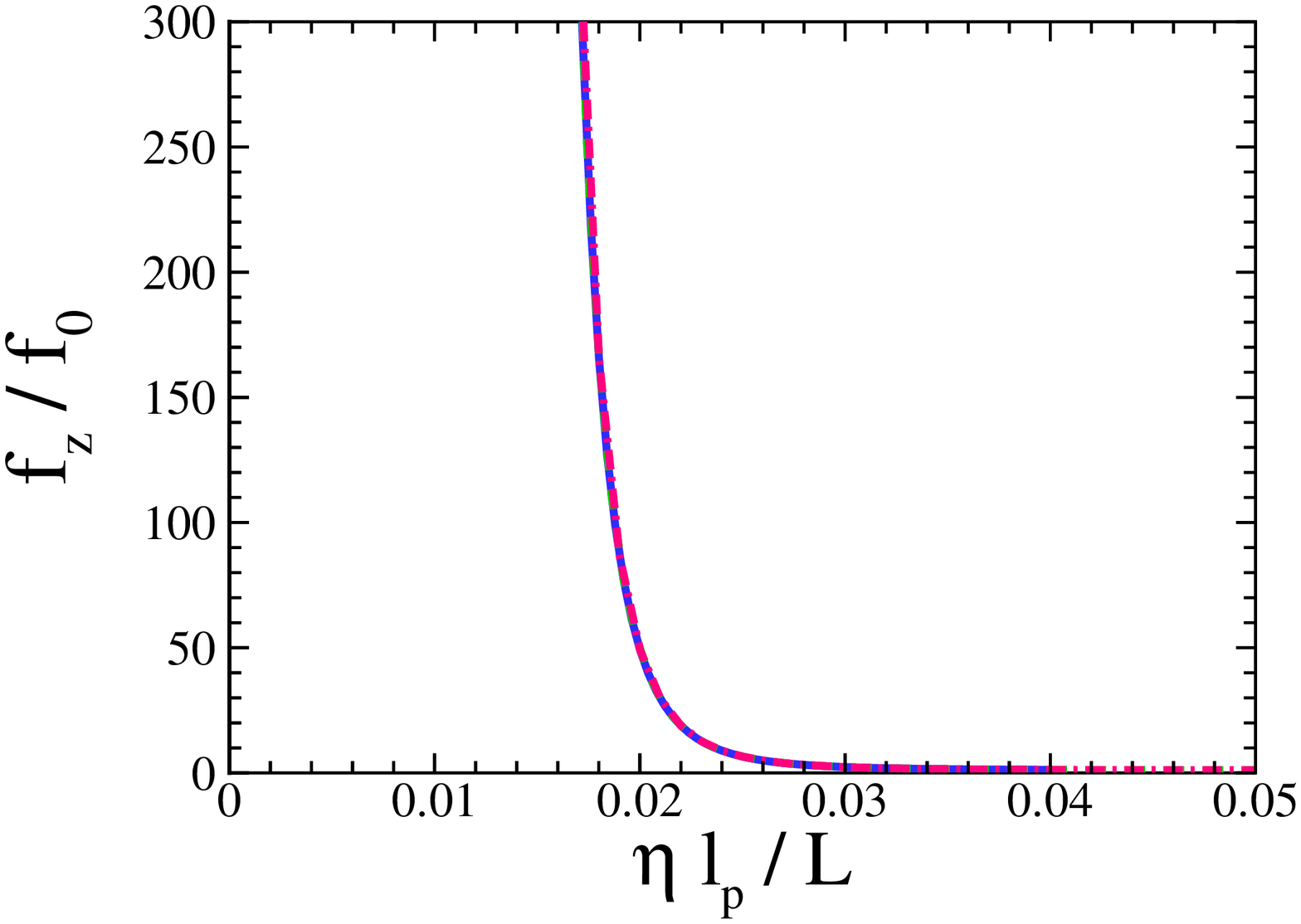}}
%\includegraphics[width=.8\columnwidth]{03-b-F(x).eps}
%\onefigure[scale=0.3]{03-b-F(x).eps}
\caption{$z$ component of the entropic force is plotted in terms of
$\eta$. To show how roughness enhances this force, we have used the corresponding force for 
a flat wall to make a non dimensional force. $f_{0}(\eta)$ is the entropic force for a wall that
is flat. As one can see, the effects due to roughness appears only at very small $\eta$. For very large value of $\eta$, that corresponds to the maximum force, the roughness has no effect. 
The parameters that we have used are: $h_0=0.1L,~qL=0.5$.
}
\label{roughfzeta}
\end{figure}

\section{Results and discussion}
Perturbation results that we have obtained in previous section allow us to explicitly
study the entropic forces and their dependence  on the parameters of our system. We first start to examine the results 
for a known case of a semiflexible polymer confined with a flat wall \cite{gholami}.
For a polymer that is confined from one of its ends with a flat wall and clamped at the other end, 
the entropic
force has only a component that is perpendicular to the wall. All lateral components, because of
symmetry, diminish to zero. 
In Fig. \ref{flatwallFzEta}, we have presented the perpendicular component of entropic force 
for a flat confining wall.
This $z$ component of the force is plotted in terms of $\eta$, the longitudinal compression of the polymer. As we have expected, for small compression of the polymer, the perpendicular force is zero. 
By increasing the compression, this force starts to have finite values. The maximum force that can be exerted by a fluctuating polymer is given by the classical Euler's buckling threshold. This maximum force can be reached at large compression $\eta$  of the polymer. Increasing the persistence length, shifts the maximum force to a higher value. 

Roughness of the wall enhances the entropic force exerted by the fluctuating polymer on the wall.
Depending on  local curvature and local slope of the wall just in front of the tip of polymer
(when it is straight and does not fluctuate), the roughness contribution will increase (or decrease, depending on the sign of curvature) the $z$ component
of the force. The slope of wall can introduce a non-zero $x$ component to the force. 
As one might expect, the roughness shows itself at very small values of $\eta$. 
For a large value of $\eta$, the tip of polymer is in contact with the wall and it has less chance 
to perform  large amplitude fluctuations and consequently this polymer will have no chance to 
detect any long range signatures of the wall that is encoded in radius of curvature. 
The effects of roughness is expected to be observed in a regime given by $\eta \ll 1/(LC)$.
In this case and to see the effects of wall roughness at small values of $\eta$, 
we have to use the  convergent formula for ${\cal Z}_0$ that was presented before.
\begin{figure}
\centering
\resizebox{0.5\textwidth}{!}{%
\includegraphics{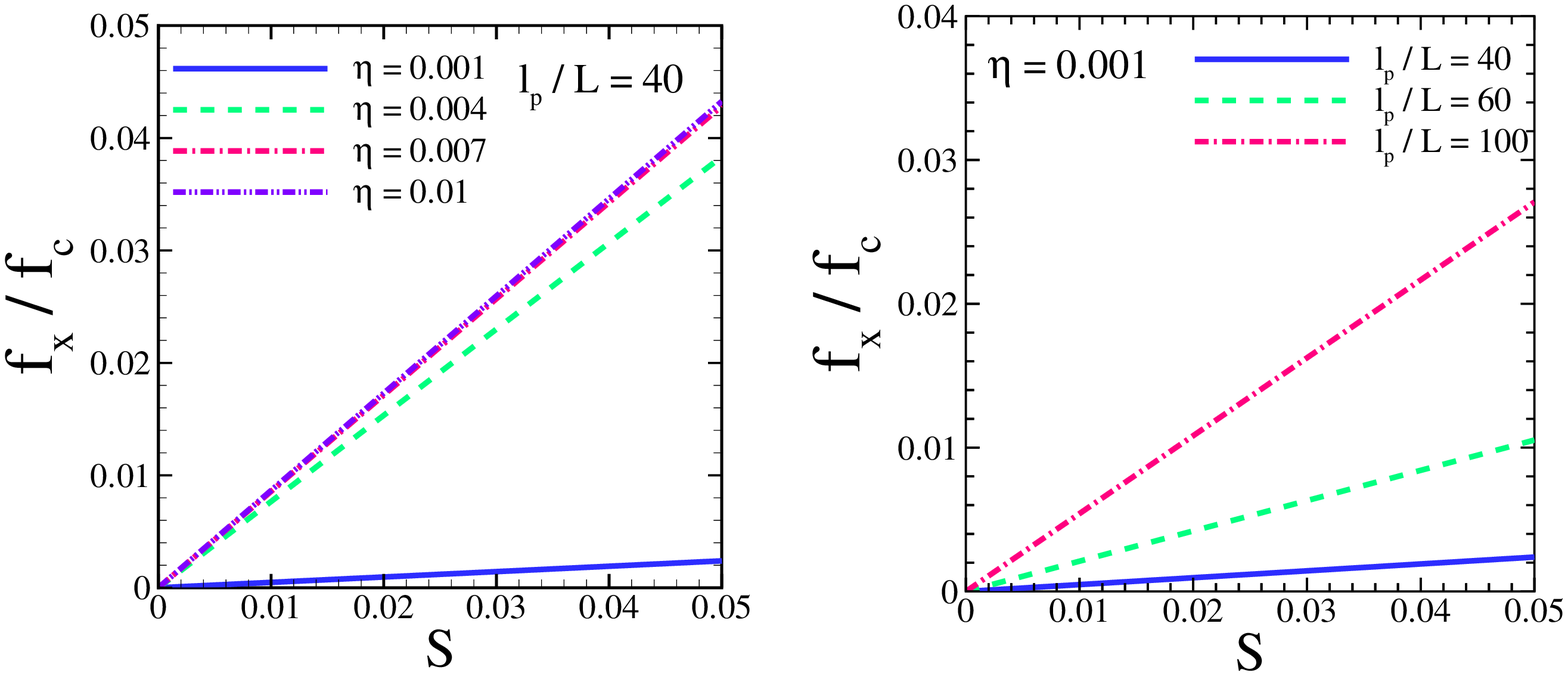}}
\caption{Lateral component of the entropic force is due to the local slope of the wall just
in front of the tip of polymer. Here $S=-h_0q\sin\phi$ measures the slope of wall. Dependence of this force to the slope is investigated for
different values of persistence length and compression $\eta$. Please note that the slope of the
wall shows itself at very low values of $\eta$.}
\label{roughfx}
\end{figure}
 
The change in perpendicular component of the force because of the roughness, is presented in 
Fig. \ref{roughfz}. The force is plotted in terms of the local curvature of the wall.  
As one can see, the force has higher values at high curvatures.
For a constant curvature, increasing the persistence length or increasing the compression $\eta$, 
will decrease the dimensionless force. In both graphs to make the force nondimensional, we have used $f_0$, the corresponding entropic force of a flat 
wall.  
In Fig. \ref{roughfzeta}, the perpendicular value of entropic force is plotted
in terms of $\eta$. As before, we have used $f_0$ to make a dimensionless  force. 
Divergence at $\eta=0$ 
is due to our the fact that $f_0$  is itself zero at this point. Again this proves that 
the curvature contribution is large at small $\eta$. At large $\eta$,
where we expect to see the maximum force, the curvature has no effect. This means that the
roughness of wall can not enhance the maximum entropic force that can be exerted on wall. 

In addition to the perpendicular entropic force, the slope of the wall just in front of the 
polymer, will introduce a lateral component to the force. As a result of 
symmetry, the force will not have any component in the $y$ direction, the direction that the wall 
is flat. In Fig. \ref{roughfx},
we have plotted the  component of the force  exerted on the wall along $x$ direction. The lateral component of the force is controlled by the local slope of the wall. 
For a constant persistence length and compression $\eta$, the force increases with increasing the 
slope $S$. For a constant slope, increasing either the persistence length or the compression $\eta$, 
will decrease the force. One should note that all of our results are the first order nontrivial corrections due to 
the roughness. As one go beyond the first order perturbation analysis, the terms that couple the  curvature and local slope of the wall are also expected to appear.   

In conclusion, we have theoretically analyzed the entropic forces exerted by a semiflexible polymer
on a curved wall. The first order nontrivial corrections for the  effects of wall roughness is investigated in details. 
In addition to the weak roughness limit, our calculations are valid for a weakly bending filament with $L/\ell_p\ll 1$. Monte Carlo simulations can be used to go beyond this approximation and achieve the results for stiff polymers.
 Two important 
complexities that we have not addressed here are related to the electrostatic and hydrodynamic effects 
which are always present in most soft matter systems.  
In poly-electrolytes, biopolymers with net electric charges, it is a known fact that the 
buckling threshold is slightly different from the Euler's classical value \cite{xabat,bucklefluc1,bucklefluc2}. In addition 
to the charging effects, the hydrodynamic effects near a rough wall may change the overall picture. 
The asymmetric hydrodynamic mobility of the monomers mediated by the wall, can modify the 
results \cite{roughhydro}. A complete theory for entropic  forces should incorporate these effects as well.     

A. N. wish to thank the Abdus-Salam international 
center for theoretical physics (ICTP), form warm hospitality during his stay.

\end{document}